\documentclass[aps,twocolumn,showkeywords,preprintnumbers,amsmath,amssymb,superscriptaddress,floatfix,nofootinbib]{revtex4}

\usepackage{lipsum}
\usepackage{graphicx}
\usepackage{epsfig}
\usepackage{epstopdf}
\usepackage{hyperref}
\usepackage{amsmath}
\usepackage{amsfonts}
\usepackage{amssymb}

\newcommand{\Slash}[1]{\ooalign{\hfil/\hfil\crcr$#1$}}

\begin{document}

\title{Role of the $\Lambda(1600)$ in the $K^-p \to \Lambda \pi^0\pi^0$ reaction}

\author{He Zhou}
\affiliation{Institute of Modern Physics, Chinese Academy of
Sciences, Lanzhou 730000, China} \affiliation{School of Nuclear
Sciences and Technology, University of Chinese Academy of Sciences,
Beijing 101408, China}

\author{Ju-Jun Xie}
\email{xiejujun@impcas.ac.cn} \affiliation{Institute of Modern
Physics, Chinese Academy of Sciences, Lanzhou 730000, China}
\affiliation{School of Nuclear Sciences and Technology, University
of Chinese Academy of Sciences, Beijing 101408, China}
\affiliation{School of Physics and Microelectronics, Zhengzhou
University, Zhengzhou, Henan 450001, China}

\begin{abstract}

Role of the $\Lambda(1600)$ is studied in the $K^- p \to \Lambda
\pi^0 \pi^0$ reaction by using the effective Lagrangian approach
near the threshold. We perform a calculation for the total and
differential cross sections by considering the contributions from
the $\Lambda(1600)$ and $\Lambda(1670)$ intermediate resonances
decaying into $\pi^0 \Sigma^{*0}(1385)$ with $\Sigma^{*0}(1385)$
decaying into $\pi^0 \Lambda$. Besides, the non-resonance process
from $u$-channel nucleon pole is also taken into account. With our
model parameters, the current experimental data on the total cross
sections of the $K^- p \to \Lambda \pi^0 \pi^0$ reaction can be well
reproduced. It is shown that we really need the contribution from
the $\Lambda(1600)$ with spin-parity $J^P = 1/2^+$, and that these
measurements can be used to determine some of the properties of the
$\Lambda(1600)$ resonance. Furthermore, we also plot the $\pi^0
\Lambda$ invariant mass distributions which could be tested by the
future experimental measurements.


\end{abstract}
\date{\today}

 \maketitle

\section{Introduction}
\label{introduction}

The $\bar{K}N$ scattering has been widely used to study the
properties of the hyperon
resonances~\cite{Khemchandani:2018amu,Sadasivan:2018jig,Feijoo:2018den,Guo:2012vv,Zhang:2013cua,Zhang:2013sva,Shi:2014vha,Zhong:2008km,Prakhov:2004an},
and it is extremely important to investigate these low excited
hyperon states through the proposed $K_L$ beam experiments at Jefferson
Lab~\cite{Amaryan:2017ldw,Zou:2016bxw}. By using a chiral unitary
approach~\cite{Magas:2005vu,Jido:2003cb,Oset:2001cn,Oset:1997it},
the meson-baryon interactions are investigated and it was found that
there are two poles in the neighbourhood of the well established
$\Lambda(1405)$ state, which is actually a superposition of these
two $J^P =1/2^-$ resonances. Recently, within a dynamical
coupled-channels model~\cite{Kamano:2014zba,Kamano:2015hxa}, some
hyperon resonance parameters are extracted through a comprehensive
partial-wave analysis of the $K^-p \to \bar{K} N$, $\pi \Sigma$,
$\pi \Lambda$, $\eta \Lambda$, and $K \Xi$ data up to invariant mass
$W = 2.1$ GeV. Among the extracted resonances, a new narrow
$\Lambda$ resonance with $J^P = 3/2^+$ is also predicted in
Refs.~\cite{Kamano:2014zba,Kamano:2015hxa}. On the contrary, Liu and
Xie~\cite{Liu:2011sw,Liu:2012ge,Liu:2012bk} analyzed the $K^- p \to
\eta \Lambda$ reaction~\cite{Starostin:2001zz} with an effective
Lagrangian approach and implied a new $\Lambda$ resonance with $J^P
= 3/2^-$. Its mass is about $1670$ MeV but its width is much small
compared with the one of the well established $\Lambda(1690)$
resonance. Thus there are still some ambiguities of the $\Lambda$
excited states needs to be clarified.

On the experimental side, the Crystal Ball Collaboration reported
the measurements with high precision of the $K^- p \to \Lambda \pi^0
\pi^0$ reaction at eight incidents of $K^-$ momenta between $514$ and
$750$ MeV, corresponding to center of mass (c.m.) energies from
$1569$ to 1676 MeV~\cite{Prakhov:2004ri}. It is shown that this
reaction is dominated by the $\pi^0 \Sigma^{*0}(1385)$ intermediate
state in $s$-channel, and the contribution of the $f_0(500)$ meson
in $t$-channel to the $K^- p \to \Lambda \pi^0 \pi^0$ reaction
appears to be very small and can be neglected. Indeed, it is shown
that the contribution of scalar meson $f_0(500)$ and $f_0(980)$ from
the $K^+ K^- \to \pi^0 \pi^0$ transition term is
negligible~\cite{Sarkar:2005ap,Xie:2015mzp}. In addition, the
strength of the total cross section of $K^- p \to \Lambda \pi^0
\pi^0$ reaction could be well reproduced in terms of the large
coupling of $\Lambda(1520)$ to $\pi\Sigma^*(1385)$, which is a
prediction of the chiral unitary
approach~\cite{Sarkar:2005ap,Roca:2006sz}. On the other hand, with
the aim for searching for the evidence for the possible $\Sigma$ excited
state with $J^P =1/2^-$, which was predicted within the unquenched
penta-quark models~\cite{Helminen:2000jb,Zhang:2004xt}, the $K^- p
\to \Lambda \pi^+ \pi^-$ reaction was investigated at the energy
region of the $\Lambda(1520)$ resonance peak by using the effective
Lagrangian approach~\cite{Wu:2009nw}, where it is found that there
is evidence for the existence of the new $\Sigma^*$ state in the
$K^- p \to \Lambda \pi^+ \pi^-$ reaction.

For the $K^- p \to \Lambda \pi^0 \pi^0$ reaction, the main
contribution is from the $\Lambda^*$ resonance through the process
$K^- p \to \Lambda^* \to \pi^0 \Sigma^{*0}(1385) \to \pi^0 \pi^0
\Lambda$. This reaction gives us a rather clean platform to study the isospin-0 $\Lambda^*$
resonances because there are no isospin-1 $\Sigma^*$ resonances that
contribute to $K^- p \to \pi^0 \Sigma^{*0}(1385)$. In the energy region of the current experimental
measurements by the Crystal Ball
Collaboration~\cite{Prakhov:2004ri}, there are two well established
$\Lambda^*$ resonances give significant contributions: the
three-star $\Lambda(1600)$ with $J^P = 1/2^+$ and the four-star
$\Lambda(1670)$ with $J^P =1/2^-$. Their Breit-Wigner masses and
widths are~\cite{Tanabashi:2018oca}:
\begin{eqnarray}
M_{\Lambda^*_1} &=& 1560 \sim 1700,~~~\Gamma_{\Lambda^*_1}
= 50 \sim 250, \\
M_{\Lambda^*_2} &=& 1660 \sim 1680,~~~\Gamma_{\Lambda^*_2} = 25 \sim
50,
\end{eqnarray}
all in units of MeV and for which we hand used the notation
$\Lambda^*_1$ and $\Lambda^*_2$ to refer to the $\Lambda(1600)$ and
$\Lambda(1670)$ resonances, respectively. It is interesting to
notice that both the mass and width of the $\Lambda(1600)$ resonance
are with large uncertainties, while the ones for $\Lambda(1670)$
resonance are much precise. Furthermore, in the work of
Ref.~\cite{Shi:2014vha}, the most precise data on the $K^- p \to
\pi^0 \Sigma^0$ reaction were analyzed to study the $\Lambda^*$
resonances, and it is found that the $\Lambda(1600)$ resonance is
definitely needed. The fitted resonance parameters for the
$\Lambda(1600)$ are $M_{\Lambda(1600)} = 1574.7 \pm 0.5$ MeV and
$\Gamma_{\Lambda(1600)} = 81.9 \pm 1.1$ MeV~\cite{Shi:2014vha}. So,
we expect that the $\Lambda(1600)$ resonance may also have a
significant contribution to the $K^- p\to \Lambda \pi^0 \pi^0$
reaction. In fact, the energy dependence of the total cross section
of $K^- p\to \Lambda \pi^0 \pi^0$ reaction~\cite{Prakhov:2004ri} has
a broad shoulder around the energy region of the $\Lambda(1600)$
state.

In the present work, based on the experimental measurements of the
Crystal Ball Collaboration~\cite{Prakhov:2004ri}, we study the role
of the $\Lambda(1600)$ and $\Lambda(1670)$ resonances in the $K^- p
\to \Lambda \pi^0 \pi^0$ reaction within the effective Lagrangian
method and the resonance model. In addition, the non-resonance
process from the $u$-channel nucleon pole is also considered as the
background. Since there are large uncertainties for the mass and
width of the $\Lambda(1600)$ resonance, we will vary them to
reproduce the experimental data.  While for the $\Lambda(1670)$
resonance, we take the average values for its mass and width as
quoted in the Particle Data Group (PDG)~\cite{Tanabashi:2018oca}.
The total and differential cross sections of the $K^- p \to \Lambda
\pi^0 \pi^0$ reaction are calculated. It is found that the
contribution of the $\Lambda(1600)$ resonance is significant, and
the experimental data on the total cross sections and angular
distributions, around the reaction energy region of the
$\Lambda(1600)$ state, can be well reproduced with the model
parameters.

The present paper is organized as follows: In
sec.~\ref{sec:formalism}, we discuss the formalism and the main
ingredients for our theoretical calculations; In
sec.~\ref{sec:results} we present our numerical results and
conclusions; A short summary is given in the last section.

\section{Formalism and ingredients} \label{sec:formalism}

The combination of the resonance model and the effective Lagrangian
approach is an important theoretical tool in describing the various
scattering processes in the resonance production
region~\cite{Cheng:2016hxi,Wu:2014yca,Xie:2013db,Xie:2014zga}. In
this section, we introduce the theoretical formalism and ingredients
to study the $K^- p \to \Lambda\pi^0 \pi^0$ reaction by using the
effective Lagrangian approach and resonance model.

\subsection{Feynman diagrams and effective interaction Lagrangian densities}

\begin{figure*}[htbp]
\centering
\includegraphics[scale=0.8]{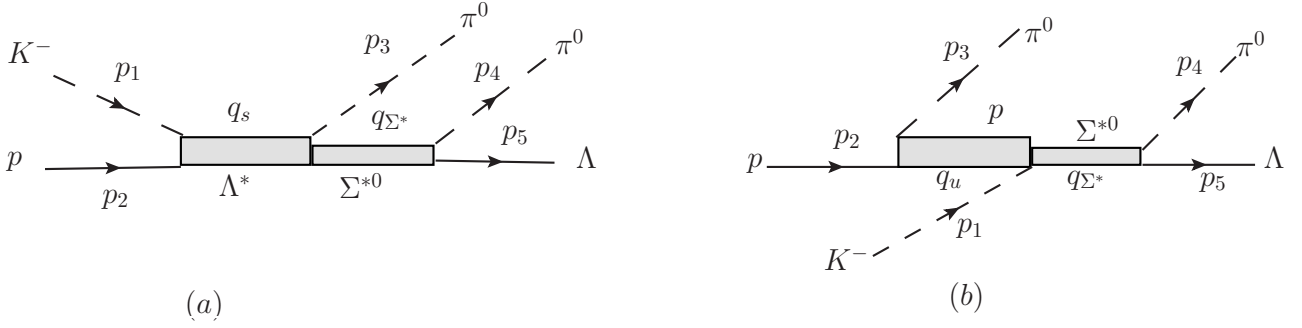}
\caption{Feynman diagrams of the $K^-p \to \pi^0\pi^0\Lambda$
reaction. The contributions from $s$-channel $\Lambda(1600)$ and
$\Lambda(1670)$ resonances and $u$-channel nucleon pole are
considered. We also show the definition of the kinematical ($p_1$,
$p_2$, $p_3$, $p_4$, $p_5$) variables that we use in the present
calculation. In addition, we use $q_s = p_1 + p_2$, $q_u = p_2 -
p_3$, and $q_{\Sigma^*} = p_4 + p_5$.} \label{fe}
\end{figure*}

The basic tree-level Feynman diagrams for the $K^- p \to \Lambda
\pi^0 \pi^0$ reaction are shown in Fig.~\ref{fe}. These include
$s$-channel $\Lambda^*$ resonances process [Fig.~\ref{fe} (a)] and
$u$-channel nucleon pole diagram [Fig.~\ref{fe} (b)]. For the $\pi^0
\Lambda$ production, we consider only the contribution from
$\Sigma^*(1385)$. The $t$-channel $K^+$ exchange term via $K^+ K^-
\to \pi^0 \pi^0$ transition is not considered since its contribution
is rather small. Besides, the $t$-channel $K^*$ exchange is also
neglected since this mechanism is much suppressed due to the highly
off-shell effect of the $K^*$ propagator when the $\pi^0 \Lambda$
invariant mass is close to the $\Sigma^*(1385)$ mass.

To evaluate the contributions of those terms shown in Fig.~\ref{fe},
the effective Lagrangian densities for relevant interaction vertexes
are needed. Following
Refs.~\cite{Doring:2010ap,Xie:2013wfa,Xie:2014kja,Xiao:2015zja,Wang:2018vlv,Wang:2017tpe,Huang:2012xj},
the Lagrangian densities used in this work are,

\begin{eqnarray}
{\cal L}_{\Lambda^*_1 \bar{K} N} &=& -\frac{g_{\Lambda^*_1 \bar{K} N}}{m_{N}+M_{\Lambda^*_1}}\bar{\Lambda}^*_1\gamma_{5}\gamma_{\mu}\partial^{\mu}\phi_{\bar K}N + {\rm h.c.}, \label{kbarnlambda1600}  \\
{\cal L}_{\Lambda^*_1 \pi \Sigma^*}&=& \frac{g_{\Lambda^*_1\pi\Sigma^*}}{m_{\pi}}\bar \Sigma^*_\mu \partial^{\mu}(\vec\tau\cdot\vec\pi)\Lambda^*_1 + {\rm h.c.}, \label{pisigmastarLambdastar1600}\\
{\cal L}_{\Lambda^*_2 \bar{K} N} &=& g_{\Lambda^*_2 \bar{K} N}\bar{\Lambda}^*_2 \phi_{\bar K} N + {\rm h.c.}, \label{kbarnlambda1670} \\
{\cal L}_{\Lambda^*_2 \pi \Sigma^*}&=& \frac{g_{\Lambda^*_2 \pi\Sigma^*}}{m_{\pi}}\bar \Sigma^*_\mu \gamma_5 \partial^{\mu}(\vec\tau\cdot\vec\pi)\Lambda^*_2 + {\rm h.c.}, \label{pisigmastarLambdastar1670} \\
{\cal L}_{\pi\Lambda\Sigma^*}&=&\frac{g_{\pi\Lambda\Sigma^*}}{m_{\pi}}\bar \Sigma^*_\mu \partial^{\mu}(\vec \tau\cdot\vec\pi)\Lambda + {\rm h.c.} , \label{pilambdasigmastar}\\
{\cal L}_{\bar{K}N\Sigma^*} &=& \frac{g_{\bar{K}N\Sigma^*}}{m_{\bar{K}}}\bar\Sigma^*_\mu\partial^{\mu}\phi_{\bar K}N + {\rm h.c.} ,\\
{\cal L}_{\pi NN}&=&-\frac{g_{\pi NN}}{2m_{N}}\bar
N\gamma_{5}\gamma_{\mu}\partial^{\mu}(\vec\tau\cdot\vec\pi) N ,
\label{pinn}
\end{eqnarray}
where $\Sigma^*_\mu$ is the Rarita-Schwinger field of the
$\Sigma^*(1385)$ resonance with spin $\frac{3}{2}$, and $\vec{\tau}$
is a usual isospin-1/2 Pauli matrix operator.

For the coupling constants in the above Lagrangian densities for
$u$-channel process, we take $g_{\pi NN}=13.45$ and
$g_{N\bar{K}\Sigma^{*}} = -3.19$ which are used in previous
works~\cite{Machleidt:1987hj,Liu:2006tf,Xie:2007qt} for studying
different processes. For the coupling constant $g_{\pi \Lambda
\Sigma^*(1385)}$ and $g_{\Lambda(1670)\bar{K}N}$, they can be
determined from the experimental observed partial decay widths of
$\Sigma^*(1385) \to \pi \Lambda$ and $\Lambda(1670) \to \bar{K}N$,
respectively.

With the effective interaction Lagrangians described by
Eqs.~\eqref{kbarnlambda1600}, \eqref{kbarnlambda1670}, and
\eqref{pilambdasigmastar}, the partial decay widths
$\Gamma_{\Sigma^* \to \pi \Lambda}$ and $\Gamma_{\Lambda^*_2 \to
\bar{K} N}$ can be easily obtained~\cite{Tanabashi:2018oca}. The
coupling constants related to the partial decay widths are written as,
\begin{eqnarray}
\Gamma_{\Lambda^*_1 \to \bar{K}N} &=&
\frac{g^2_{\Lambda^*_1\bar{K}N}}{2\pi}(E_N - m_N)
\frac{p_{\bar{K}N}}{M_{\Lambda^*_1}},
\\
\Gamma_{\Lambda^*_2 \to \bar{K}N} &=&
\frac{g^2_{\Lambda^*_2\bar{K}N}}{2\pi}(E_N + m_N) \frac{p_{\bar{K}N}}{M_{\Lambda^*_2}}, \\
\Gamma_{\Sigma^* \to \pi \Lambda} &=& \frac{g^2_{\Sigma^* \pi
\Lambda}}{12\pi}(E_\Lambda + m_{\Lambda})
\frac{p^3_{\pi\Lambda}}{m^2_{\pi}m_{\Sigma^*}},
\end{eqnarray}
with
\begin{eqnarray}
E_N &=& \frac{M^2_{\Lambda^*_1/\Lambda^*_2} + m^2_N - m^2_{\bar{K}}}{2M_{\Lambda^*_1/\Lambda^*_2}}, \\
p_{\bar{K} N} &=& \sqrt{E^2_N - m^2_N}, \\
E_{\Lambda} &=& \frac{m^2_{\Sigma^*} + m^2_\Lambda - m^2_\pi}{2m_{\Sigma^*}} , \\
p_{\pi \Lambda} &=& \sqrt{E^2_\Lambda - m^2_\Lambda}.
\end{eqnarray}
With the masses, widths and branching ratios of $\Lambda(1670)$ and
$\Sigma^*(1385)$ resonances quoting in
PDG~\cite{Tanabashi:2018oca}, the numerical results for the relevant
coupling constants are listed in Table~\ref{table:tab1}, while the
other coupling constants needed in this work will be discussed
below.

\begin{table}[htbp]
\caption{Relevant parameters used in the present calculation. The
masses, widths and branching ratios of $\Lambda(1670)$ and
$\Sigma^*(1385)$ resonances are taken from
PDG~\cite{Tanabashi:2018oca}, while for the $\Lambda(1600)$
resonance, these values are determined to the experimental data.}
\label{table:tab1}
\begin{tabular}{ c c c c c c c }
\hline \hline  State ($J^P$) & Mass & Width  & Decay  & Branching & $g^2/4\pi$ \\
 & (MeV) & (MeV)  & mode & ratio ($\%$) & \\
\hline
$\Sigma^{*}(1385)$ ($\frac{3}{2}^+$) & 1385   & 37    & $\pi \Lambda$       & 87   & 0.12     \\
$\Lambda(1670)$ ($\frac{1}{2}^-$)    & 1670   & 35    & $\bar{K} N$         & 25   & 0.009    \\
                                     &        &       & $\pi\Sigma^*(1385)$ & 22.4 & 1.07     \\
$\Lambda(1600)$ ($\frac{1}{2}^+$)    & 1580   & 150   & $\bar{K} N$         & 22.5 & 1.56     \\
                                     &        &       & $\pi\Sigma^*(1385)$ & 6.5  & 0.05     \\
\hline \hline
\end{tabular}
\end{table}

\subsection{Propagators and form factors}

To get the scattering amplitude of the $K^- p \to \Lambda \pi^0
\pi^0$ reaction corresponding to the Feynman diagrams shown in
Fig.~\ref{fe}, we also need the propagators for spin $\frac{1}{2}$
particles: nucleon, $\Lambda(1600)$ and $\Lambda(1670)$, and
$\Sigma^*(1385)$ resonance with spin $\frac{3}{2}$,
\begin{eqnarray}
G_p(q_u) &=& i \frac{ \Slash{q}_u + m_N}{q_u^2-m^2_p}, \\
G_{\Lambda^*_1/\Lambda^*_2}(q_s) &=& i \frac{\Slash{q}_s + M_{\Lambda^*_1/\Lambda^*_2}}{q_s^2-M^2_{\Lambda^*_1/\Lambda^*_2} + i M_{\Lambda^*_1/\Lambda^*_2} \Gamma_{\Lambda^*_1/\Lambda^*_2}}, \\
G^{\mu\nu}_{\Sigma^*}(q_{\Sigma^*}) &=& i
\frac{(\Slash{q}_{\Sigma^*} +
m_{\Sigma^*})P^{\mu\nu}(q_{\Sigma^*})}{q^2_{\Sigma^*} -
m^2_{\Sigma^*} + im_{\Sigma^*} \Gamma_{\Sigma^*}} ,
\end{eqnarray}
with
\begin{eqnarray}
P^{\mu\nu}(q_{\Sigma^*}) &=& -g^{\mu\nu} +
\frac{1}{3}\gamma^{\mu}\gamma^{\nu} + \frac{2}{3}
\frac{q^{\mu}_{\Sigma^*} q^{\nu}_{\Sigma^*}}{m^2_{\Sigma^*}}
\nonumber \\
&& + \frac{1}{3m_{\Sigma^*}}(\gamma^{\mu}q^{\nu}_{\Sigma^*} -
\gamma^{\nu}q^{\mu}_{\Sigma^*}),
\end{eqnarray}
where $q_u$, $q_s$ and $q_{\Sigma^*}$ are the momenta of nucleon
pole in $u$-channel, $\Lambda(1600)$ or $\Lambda(1670)$ resonance in
$s$-channel, and $\Sigma^*(1385)$ resonance, respectively.

Finally, we need to also include the off-shell form factors in the
scattering amplitudes. There is no unique theoretical way to
introduce the form factors, hence, we adopt here the common scheme
used in many previous works~\cite{Liu:2006tf,Xie:2007qt,Xie:2010yk},

\begin{eqnarray}
&& f_i =\frac{\Lambda^4_i}{\Lambda^4_i+(q_i^2-M_i^2)^2},
\quad i= s, u, \Sigma^* \\
&&\quad {\rm with} ~~~~~~ \left\{\begin{array}{l}  q_s^2=s, \,
q_u^2= u, \, q_{\Sigma^*}^2= M^2_{\pi^0 \Lambda} \cr M_u = m_N, \,
M_{\Sigma^*} = m_{\Sigma^*}, \cr M_s = M_{\Lambda^*_1/\Lambda^*_2},
,
\end{array}\right. \label{formfactor}
\end{eqnarray}
where $s$ and $u$ are the Lorentz-invariant Mandelstam variables,
while $M_{\pi^0 \Lambda}$ is the invariant mass of the $\pi^0
\Lambda$ system. In the present calculation, $q_s= p_1 + p_2$, $q_u
= p_2-p_3$, and $q_{\Sigma^*} = p_4 + p_5$ are the 4-momenta of
intermediate $\Lambda(1600)$ or $\Lambda(1670)$ resonance, exchanged
nucleon pole in the $u$-channel, and the $\Sigma^*(1385)$ resonance
decaying into $\pi^0 \Lambda$, respectively, while $p_1$, $p_2$,
$p_3$, $p_4$, and $p_5$ are the 4-momenta for $K^-$, $p$, $\pi^0$,
$\pi^0$, and $\Lambda$, respectively. Besides, we will consider same
cut-off values for the background and resonant terms, i.e.
$\Lambda_s  = \Lambda_u$. Note that the numerical results are not
sensitive to $\Lambda_s$ and $\Lambda_{\Sigma^*}$.

\subsection{Scattering amplitudes}

With the effective interaction Lagrangian densities given above, we
can easily construct the invariant scattering amplitudes for the
$K^- p \to \Lambda \pi^0 \pi^0$ reaction corresponding to the
diagrams shown in Fig.~\ref{fe}:
\begin{equation}
\mathcal{M}=\mathcal{M}(\Lambda^*_1) + \mathcal{M}(\Lambda^*_2) +
\mathcal{M}(N).
\end{equation}

Each of the above amplitudes can be obtained straightforwardly as,
\begin{eqnarray}
\mathcal{M}(i) = \bar{u}(p_5,s_{\Lambda})G^{\mu\nu}_{\Sigma^*}{\cal
A}_{\mu\nu}(i) u(p_2,s_p),
\end{eqnarray}
where $s_{\Lambda}$ and $s_p$ are the spin polarization variables
for the final $\Lambda$ and initial proton, respectively. The
reduced $A^{\mu\nu}(i)$ can be also easily obtained:
\begin{eqnarray}
{\cal A}^{\mu\nu}(\Lambda^*_1) &=& -i g_1 p^\mu_4 p^{\nu}_{3}
G_{\Lambda^*_1}(q_s)
\gamma_{5} \Slash{p}_1 f_s(\Lambda^*_1) f_{\Sigma^*},  \\
{\cal A}^{\mu\nu}(\Lambda^*_2) &=&
 g_2 p^\mu_4 p^{\nu}_{3} \gamma_{5} G_{\Lambda^*_2}(q_s) f_s(\Lambda^*_2) f_{\Sigma^*}, \\
{\cal A}^{\mu\nu}(N) &=& -i g_3 p^\mu_4 p^{\nu}_{1} G_N(q_u)
\gamma_{5} \Slash{p}_3 f_u f_{\Sigma^*},
\end{eqnarray}
with
\begin{eqnarray}
g_1 &=&  \frac{g_{\Sigma^*\pi\Lambda}
g_{\Lambda^*_1\pi\Sigma^*}g_{\Lambda^*_1\bar{K}N}}{m^2_{\pi}(m_{N} +
M_{\Lambda^*_1})}, \\
g_2 &=& \frac{g_{\Sigma^*\pi\Lambda}g_{\Lambda^*_2\pi\Sigma^*}g_{\Lambda^*_2\bar{K}N}}{m^2_{\pi}}, \\
g_3 &=& \frac{g_{\Sigma^*\pi\Lambda} g_{\Sigma^* \bar{K}N}g_{\pi
NN}}{2m_{\pi}m_{\bar{K}}m_N}.
\end{eqnarray}

Then, the cross section for the $K^- p \to \Lambda \pi^0 \pi^0$
reaction can be calculated
by~\cite{Tanabashi:2018oca,Xie:2015zga}~\footnote{Note that the
total squared amplitude for $K^- p \to \pi^0 \pi^0 \Lambda$ reaction
is symmetrized in the momenta $p_3$ and $p_4$ to account for the two
$\pi^0$ in the final state.}
\begin{eqnarray}
d\sigma & = & \frac{1}{4}\frac{1}{(2\pi)^5}\frac{m_{p}}{\sqrt{(p_{1}
\cdot p_{2})^2 - m_{p}^2 m^2_{K^-}}} \times \sum_{s_p,s_{\Lambda}}
|\mathcal{M}|^2 \nonumber \\
&& \times \frac{d^3p_{3}}{2E_{3}}\frac{d^3p_{4}}{2E_{4}}
\frac{m_{\Lambda}d^3p_{5}}{E_{5}}\delta^4(p_{1} + p_{2} - p_3 -
p_{4} - p_{5}) \nonumber \\
&=& \frac{1}{2^{10}\pi^5} \frac{m_{p}m_{\Lambda}}{\sqrt{s[(p_1 \cdot
p_{2})^2 - m^2_{p} m^2_{K^-}]}} \sum_{s_p,s_{\Lambda}}
|\mathcal{M}|^2
\\ && \times |\vec p_{3}| |\vec p_5^{~*}|
dM_{\pi^0\Lambda}d\Omega_{3} d\Omega^*_{5},
\end{eqnarray}
with $s = (p_1 + p_2)^2 = m^2_p + m^2_{K^-} + 2 p_1 \cdot p_2$, and
$\vec p_5^{~*}$ and $\Omega^*_5$ are the three-momentum and solid
angle of the out going $\Lambda$ in the center-of-mass (c.m.) frame
of the final $\pi^0 \Lambda$ system, while $\vec p_{3}$ and
$\Omega_3$ are the three-momentum and solid angle of the $\pi^0$
meson in the c.m. frame of the initial $K^- p$ system. Note that we
have already taken into account the factor $1/2$ for the identity
for the two pions in the final state.

\section{Numerical results and discussions}   \label{sec:results}

\begin{figure}[htbp]
\centering
\includegraphics[scale=0.33]{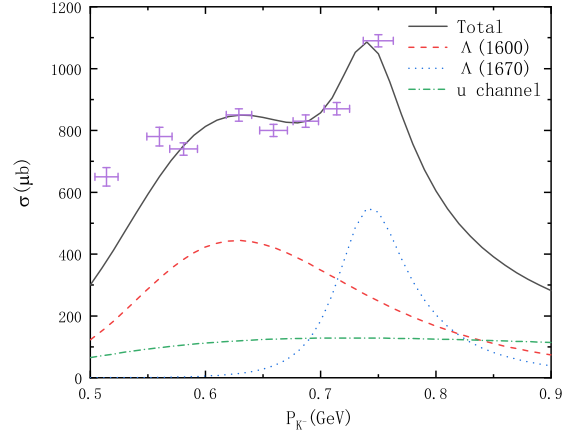}
\caption{Theoretical results of the total cross sections of $K^- p
\to \Lambda \pi^0\pi^0$ reaction. The experimental data are taken
from Ref.~\cite{Prakhov:2004ri}.} \label{fig:tcs}
\end{figure}

The theoretical results for the total cross sections for beam
momenta $p_{K^-}$ (module of the three momentum $\vec{p}_1$) from
$0.5$ to $0.9$ GeV are shown in Fig.~\ref{fig:tcs}, where we have
investigated the role of $\Lambda(1600)$, $\Lambda(1670)$ and the
$u$-channel process in describing the total cross sections. The
contributions from different mechanisms are shown separately. The
red dashed, blue dotted, and green dash-dotted curves stand for
contributions from the $\Lambda(1600)$, $\Lambda(1670)$ and
$u$-channel, respectively. Their total contributions are shown by
the solid line. The theoretical numerical results are obtained with
the following parameters: $\Lambda_s$ = 600 MeV for the
$\Lambda(1600)$ and $\Lambda(1670)$ resonances, $\Lambda_u =
\Lambda_{\Sigma^*}$ = 600 MeV, $M_{\Lambda^*_1} = 1580$ MeV,
$\Gamma_{\Lambda^*_1} = 150$ MeV, $g_{\Lambda^*_1 \pi \Sigma^*} =
0.79$, and $g_{\Lambda^*_2 \pi \Sigma^*} = 3.67$.

From Fig.~\ref{fig:tcs}, one can see that we can fairly well
reproduce the experimental data of Ref.~\cite{Prakhov:2004ri}, and
that the $\Lambda(1600)$ resonance gives a dominant contribution to
the reaction around $p_{K^-} = 630$ MeV, while the contribution of
$\Lambda(1670)$ is significant around $p_{K^-} = 750$ MeV. On the
other hand, it is seen clearly that the inclusion of the
$\Lambda(1600)$ resonance is crucial to achieve a fairly good
description of the experimental data. However, we can not describe
the enhancement at low energy region, where it could be explained by
the tail of the contribution of the $\Lambda(1520)$ in
Refs.~\cite{Sarkar:2005ap,Roca:2006sz}, and it may also be explained
by the possible $\Sigma^*(1380) \to \pi \Lambda$ in $s$ wave as
proposed in Ref.~\cite{Wu:2009nw}. Such calculations are beyond the
scope of the present investigation but we will clarify this issue in
a future study.

\begin{figure}[htbp]
\centering
\includegraphics[scale=0.27]{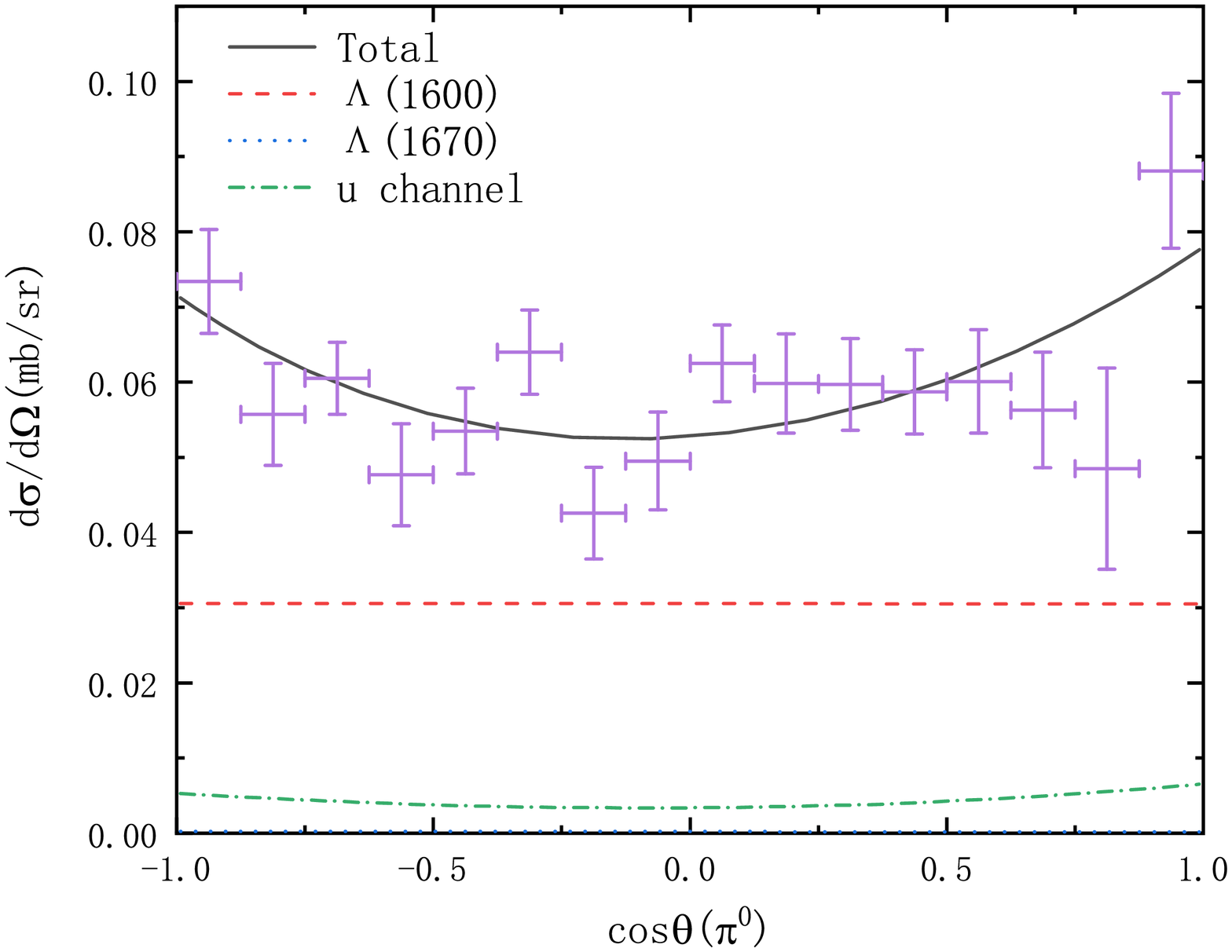}
\includegraphics[scale=0.27]{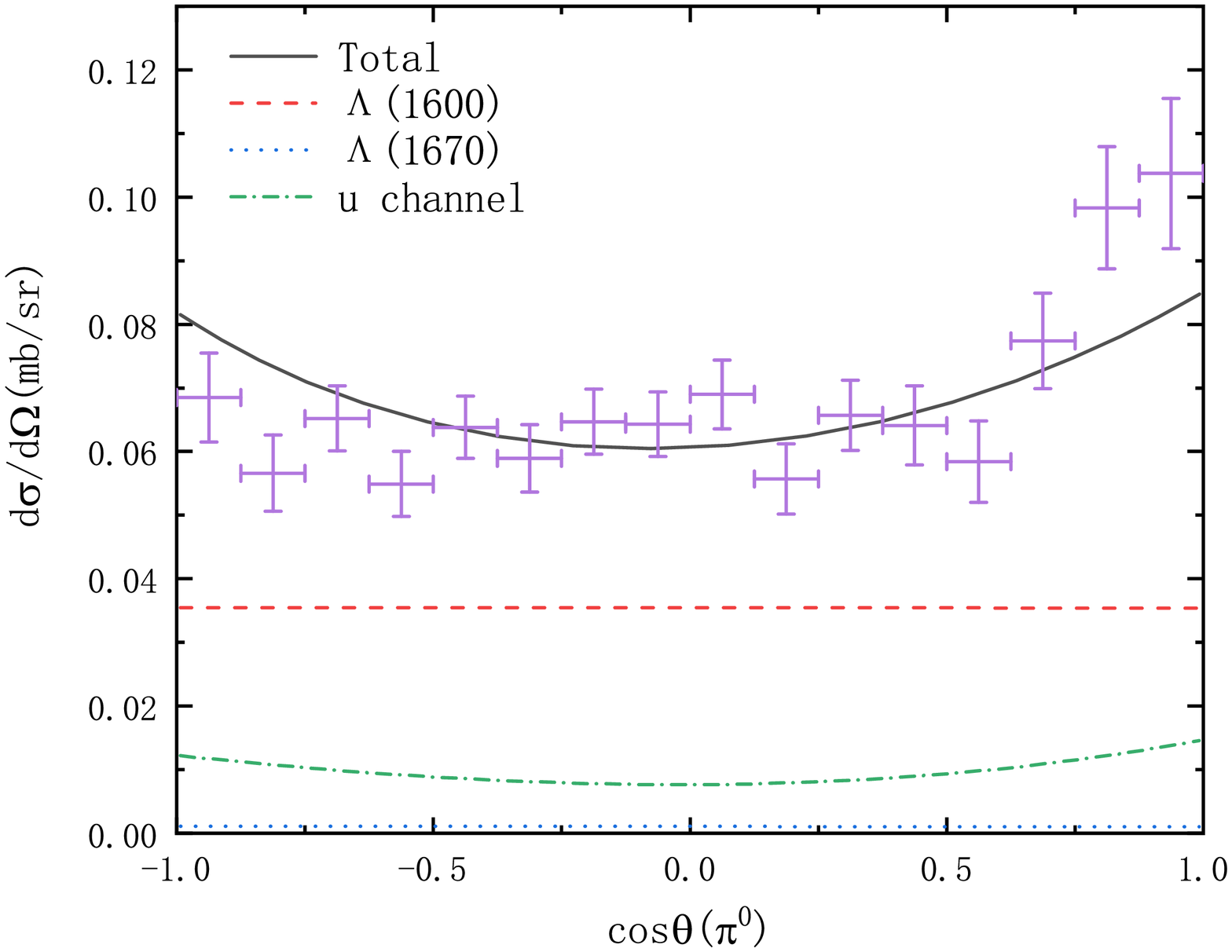}
\includegraphics[scale=0.27]{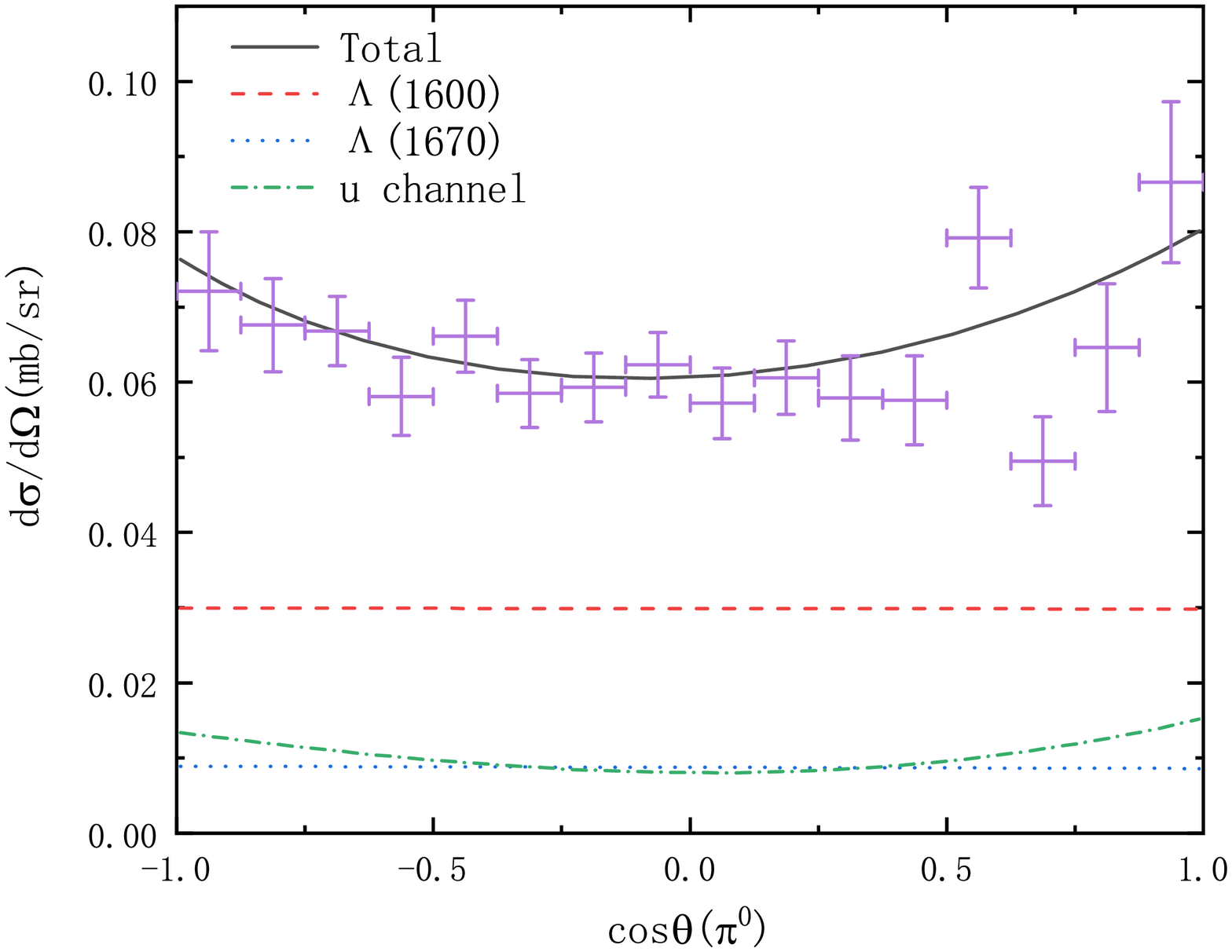}
\caption{Angular differential cross sections for the $K^-p \to
\Lambda \pi^0 \pi^0$ reaction as a function of ${\rm cos}\theta$
with $\theta$ the angel between the $\pi^0$ direction and the beam
direction in the overall c.m. system at $p_{K^-}$ = 581 (up), 629
(middle), and 687 MeV (down). The experimental data are taken from
Ref.~\cite{Prakhov:2004ri}.} \label{angledistribution}
\end{figure}

With the obtained strong coupling constants $g_{\Lambda^*_1 \pi
\Sigma^*}$ and $g_{\Lambda^*_2 \pi \Sigma^*}$, we have evaluated the
$\Lambda(1600)$ and $\Lambda(1670)$ resonances to the $\pi
\Sigma^*(1385)$ partial decay width:
\begin{eqnarray}
\Gamma_{\Lambda^*_1/\Lambda^*_2 \to \pi \Sigma^*} &=&
\frac{g^2_{\Lambda^*_1/\Lambda^*_2 \pi
\Sigma^*}M_{\Lambda^*_1/\Lambda^*_2}}{6\pi
m^2_{\pi}m_{\Sigma^*}}(E_{\Sigma^*} \pm m_{\Sigma^*})
p^3_{\pi\Sigma^*}, \nonumber
\end{eqnarray}
with
\begin{eqnarray}
E_{\Sigma^*} &=& \frac{M^2_{\Lambda^*_1/\Lambda^*_2} + m^2_{\Sigma^*} - m^2_\pi}{2M_{\Lambda^*_1/\Lambda^*_2}} , \\
p_{\pi \Sigma^*} &=& \sqrt{E^2_{\Sigma^*} - m^2_{\Sigma^*}},
\end{eqnarray}
as deduced from the Lagrangians of
Eq.~\eqref{pisigmastarLambdastar1600} and
Eq.~\eqref{pisigmastarLambdastar1670}. With the partial decay
widths, we can then obtain the branching ratios. The numerical
predictions for these branching ratios are also given in
Table~\ref{table:tab1}. Note that the uncertainties of the coupling
constants and cut off parameters are not studied in this work,
since, including such effects, the scattering amplitudes would be
more complex due to additional model parameters, and we cannot
exactly determine these parameters. Thus, we leave these
investigations to further studies when more precise experimental
measurements become available.

In addition to the total cross sections, we also compute the angle
distributions for $K^- p \to \Lambda \pi^0 \pi^0$ reaction. The
corresponding theoretically numerical results at $p_{K^-}$ = 581,
629, and 687 MeV, where the contribution of the $\Lambda(1600)$
resonance is dominant, are shown in Fig.~\ref{angledistribution}.
For comparison, we also show the experimental data from
Ref.~\cite{Prakhov:2004ri}. It is obvious that we can fairly well
reproduce the current experimental data on the angular distribution
of the $K^- p \to \Lambda \pi^0 \pi^0$ reaction thanks to the
contribution of the $\Lambda(1600)$ resonance.

\begin{figure}[htbp]
\centering
\includegraphics[scale=0.27]{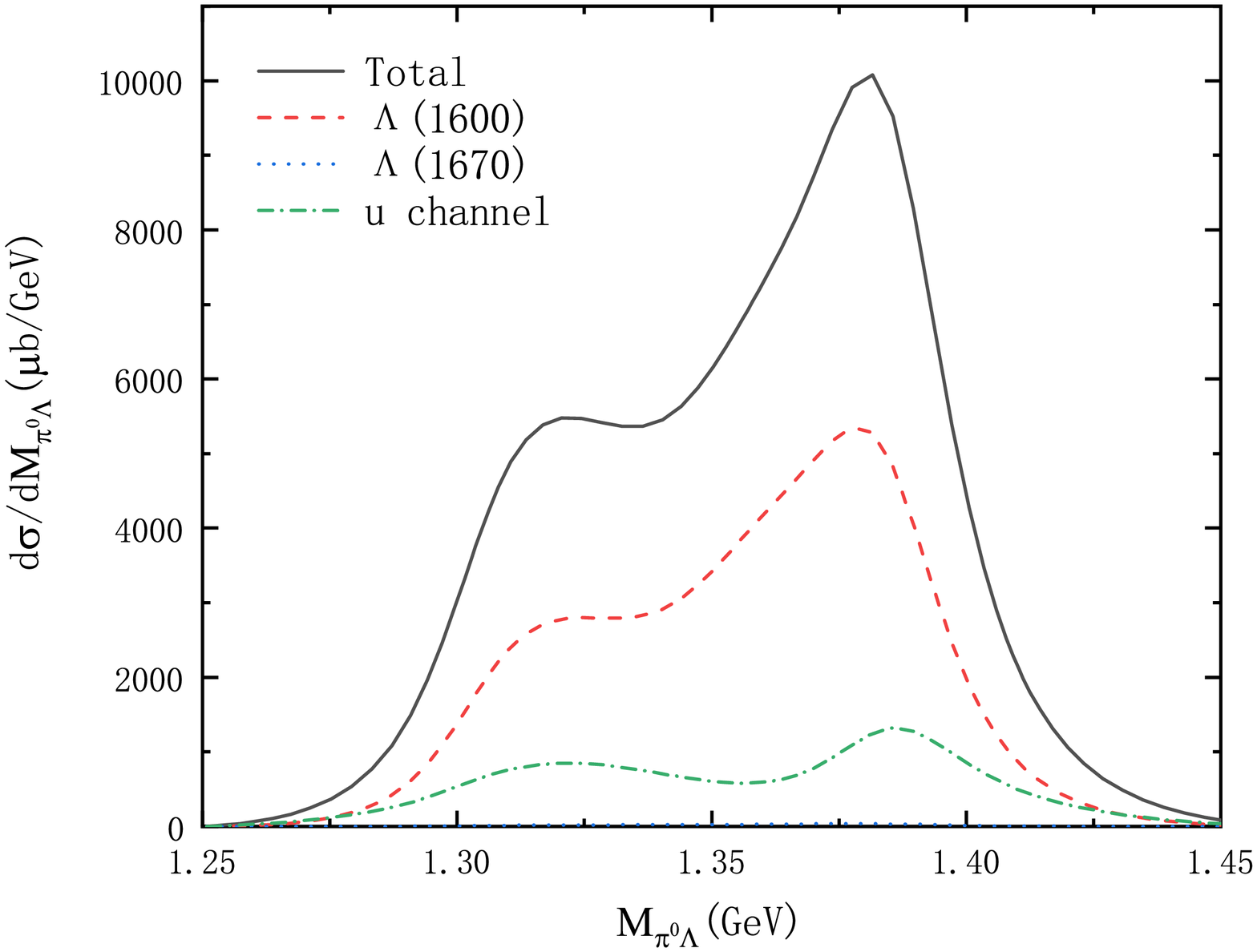}
\includegraphics[scale=0.27]{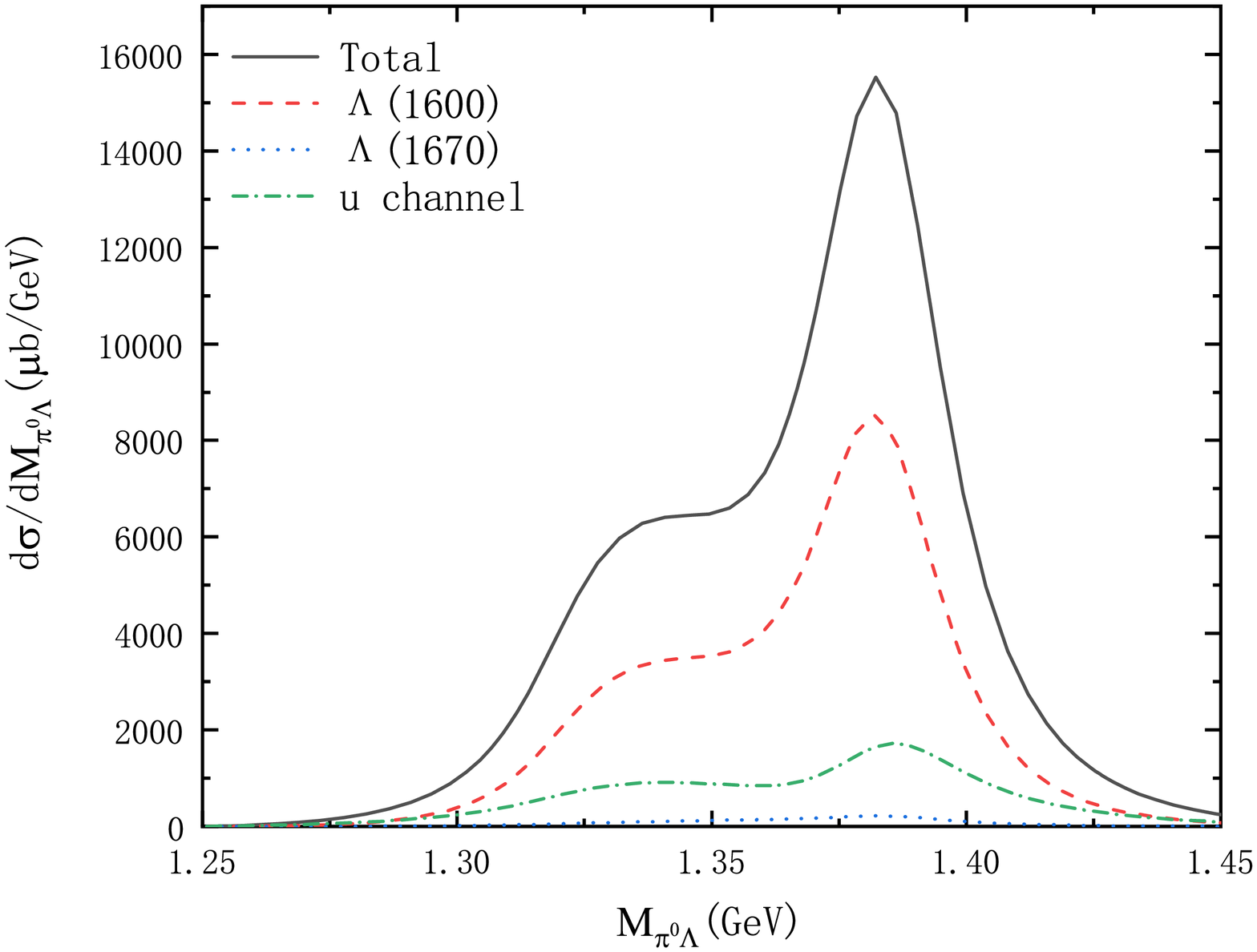}
\includegraphics[scale=0.27]{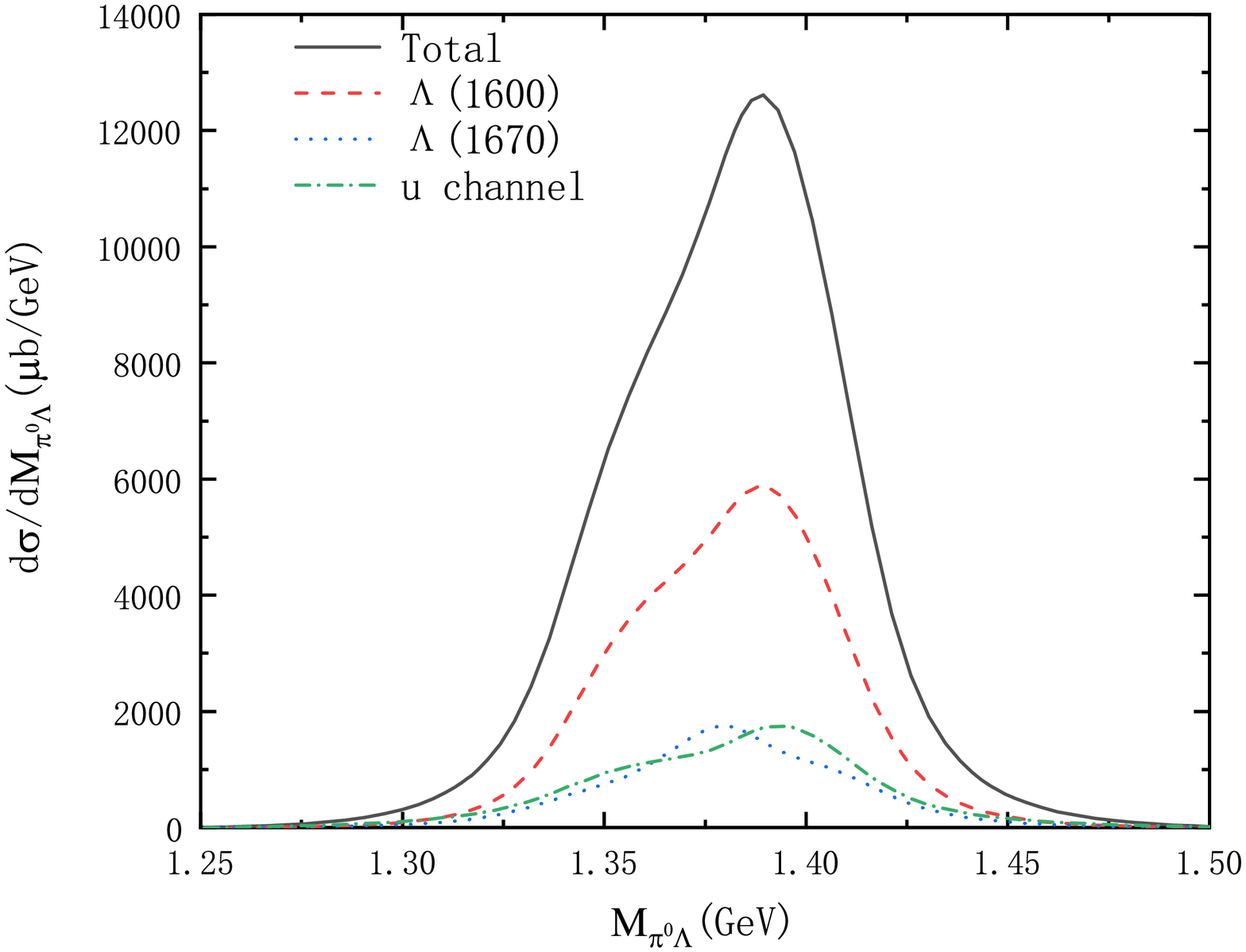}
\caption{The $\pi^0 \Lambda$ invariant mass distribution of $K^- p
\to \Lambda \pi^0 \pi^0$ reaction at $p_{K^-}$ = 581 (up), 629
(middle), and 687 MeV (down).} \label{invariantmassdistribution}
\end{figure}

Finally, in Fig.~\ref{invariantmassdistribution}, we show the
theoretical results on the differential cross section $d\sigma
\slash dM_{\pi^0\Lambda}$ as a function of the invariant mass of a
pair of $\pi^0\Lambda$ for the values of $K^-$ momentum, $581$, 629
and 687 MeV. From these figures, we see that the shape of the $\pi^0
\Lambda$ invariant mass distributions are different with the beam
energy increasing. We hope that the future experimental measurements
can check our model calculations.

\section{Summary}

In summary, we have investigated the total and differential cross
sections of the $K^- p \to \Lambda \pi^0 \pi^0$ reaction within an
effective Lagrangian approach and the resonance model. The role
played by the $\Lambda(1600)$ and $\Lambda(1670)$ resonances are
studied. It is shown that our model calculations lead to a fair
description of the experimental data on the total cross section
except for the low energy date. The scheme proposed herein should be
supplemented with some other reaction mechanisms which could improve
the achieved description of the low energy enhancement. Indeed, as
is proposed in Refs.~\cite{Sarkar:2005ap,Roca:2006sz} the
$\Lambda(1520)$ plays an important role in the $K^- p \to \Lambda
\pi^0 \pi^0$ reaction with the $K^- p \to \pi \Sigma^*(1385)$
amplitude obtained from the chiral unitary approach. However, we
have shown here that the $\Lambda(1600)$ and $\Lambda(1670)$
resonances give dominant contributions, and the consideration of the
$\Lambda(1600)$ resonance is crucial.

Finally, we would like to stress that, thanks to the important role
played by the resonant contribution of $\Lambda(1600)$ resonance in
the $K^- p \to \Lambda \pi^0 \pi^0$ reaction, we can describe
experimental data on the total cross section and angle
distributions. Accurate data for this reaction can be used to
improve our knowledge of some $\Lambda(1600)$ properties, which are
at present poorly known. This work constitutes a first step in this
direction.

\section*{Acknowledgements}

One of us (He Zhou) would like to thank Xu Zhang for useful
discussions. This work is partly supported by the National Natural
Science Foundation of China under Grant Nos. 11735003, 1191101015
and 11475227, and by the Youth Innovation Promotion Association CAS
(No.~2016367).


\begin{thebibliography}{99}

\bibitem{Khemchandani:2018amu}
  K.~P.~Khemchandani, A.~Mart\'inez Torres and J.~A.~Oller,
  Phys.\ Rev.\ C {\bf 100}, 015208 (2019).

\bibitem{Sadasivan:2018jig}
  D.~Sadasivan, M.~Mai and M.~D\"oring,
  Phys.\ Lett.\ B {\bf 789}, 329 (2019).

\bibitem{Feijoo:2018den}
  A.~Feijoo, V.~Magas and A.~Ramos,
  Phys.\ Rev.\ C {\bf 99}, 035211 (2019).

\bibitem{Guo:2012vv}
  Z.~H.~Guo and J.~A.~Oller,
  Phys.\ Rev.\ C {\bf 87}, 035202 (2013).

\bibitem{Zhang:2013cua}
  H.~Zhang, J.~Tulpan, M.~Shrestha and D.~M.~Manley,
  Phys.\ Rev.\ C {\bf 88}, 035204 (2013).

\bibitem{Zhang:2013sva}
  H.~Zhang, J.~Tulpan, M.~Shrestha and D.~M.~Manley,
  Phys.\ Rev.\ C {\bf 88}, 035205 (2013).

\bibitem{Shi:2014vha}
  J.~Shi and B.~S.~Zou,
  Phys.\ Rev.\ C {\bf 91}, 035202 (2015).

\bibitem{Zhong:2008km}
  X.~H.~Zhong and Q.~Zhao,
  Phys.\ Rev.\ C {\bf 79}, 045202 (2009).

\bibitem{Prakhov:2004an}
  S.~Prakhov {\it et al.} [Crystall Ball Collaboration],
  Phys.\ Rev.\ C {\bf 70}, 034605 (2004).


\bibitem{Amaryan:2017ldw}
  S.~Adhikari {\it et al.} [GlueX Collaboration],
  arXiv:1707.05284 [hep-ex].

\bibitem{Zou:2016bxw}
  B.~S.~Zou,
  arXiv:1603.03927 [hep-ph].

\bibitem{Magas:2005vu}
  V.~K.~Magas, E.~Oset and A.~Ramos,
  Phys.\ Rev.\ Lett.\  {\bf 95}, 052301 (2005).

\bibitem{Jido:2003cb}
  D.~Jido, J.~A.~Oller, E.~Oset, A.~Ramos and U.~G.~Meissner,
  Nucl.\ Phys.\ A {\bf 725}, 181 (2003).

\bibitem{Oset:2001cn}
  E.~Oset, A.~Ramos and C.~Bennhold,
  Phys.\ Lett.\ B {\bf 527}, 99 (2002) Erratum: [Phys.\ Lett.\ B {\bf 530}, 260 (2002)]

\bibitem{Oset:1997it}
  E.~Oset and A.~Ramos,
  Nucl.\ Phys.\ A {\bf 635}, 99 (1998).

\bibitem{Kamano:2014zba}
  H.~Kamano, S.~X.~Nakamura, T.-S.~H.~Lee and T.~Sato,
  Phys.\ Rev.\ C {\bf 90}, 065204 (2014).

\bibitem{Kamano:2015hxa}
  H.~Kamano, S.~X.~Nakamura, T.-S.~H.~Lee and T.~Sato,
  Phys.\ Rev.\ C {\bf 92}, 025205 (2015)
  Erratum: [Phys.\ Rev.\ C {\bf 95}, 049903 (2017)].

\bibitem{Liu:2011sw}
  B.~C.~Liu and J.~J.~Xie,
  Phys.\ Rev.\ C {\bf 85}, 038201 (2012).

\bibitem{Liu:2012ge}
  B.~C.~Liu and J.~J.~Xie,
  Phys.\ Rev.\ C {\bf 86}, 055202 (2012).

\bibitem{Liu:2012bk}
  B.~C.~Liu and J.~J.~Xie,
  Few Body Syst.\  {\bf 54}, 1131 (2013).

\bibitem{Starostin:2001zz}
  A.~Starostin {\it et al.} [Crystal Ball Collaboration],
  Phys.\ Rev.\ C {\bf 64}, 055205 (2001).

\bibitem{Prakhov:2004ri}
  S.~Prakhov {\it et al.},
  Phys.\ Rev.\ C {\bf 69}, 042202 (2004).


\bibitem{Xie:2015mzp}
  J.~J.~Xie, W.~H.~Liang and E.~Oset,
  Phys.\ Rev.\ C {\bf 93}, 035206 (2016).

\bibitem{Sarkar:2005ap}
  S.~Sarkar, E.~Oset and M.~J.~Vicente Vacas,
  Phys.\ Rev.\ C {\bf 72}, 015206 (2005).

\bibitem{Roca:2006sz}
  L.~Roca, S.~Sarkar, V.~K.~Magas and E.~Oset,
  Phys.\ Rev.\ C {\bf 73}, 045208 (2006).

\bibitem{Helminen:2000jb}
  C.~Helminen and D.~O.~Riska,
  Nucl.\ Phys.\ A {\bf 699}, 624 (2002).

\bibitem{Zhang:2004xt}
  A.~Zhang, Y.~R.~Liu, P.~Z.~Huang, W.~Z.~Deng, X.~L.~Chen and S.~L.~Zhu,
  High Energy Phys. Nucl. Phys. {\bf 29}, 250 (2005).

\bibitem{Wu:2009nw}
  J.~J.~Wu, S.~Dulat and B.~S.~Zou,
  Phys.\ Rev.\ C {\bf 81}, 045210 (2010).

\bibitem{Tanabashi:2018oca}
  M.~Tanabashi {\it et al.} [Particle Data Group],
  Phys.\ Rev.\ D {\bf 98}, 030001 (2018).

\bibitem{Cheng:2016hxi}
  C.~Cheng, J.~J.~Xie and X.~Cao,
  Commun.\ Theor.\ Phys.\  {\bf 66}, 675 (2016).

\bibitem{Wu:2014yca}
  C.~Z.~Wu, Q.~F.~L\"u, J.~J.~Xie and X.~R.~Chen,
  Commun.\ Theor.\ Phys.\  {\bf 63}, 215 (2015).

\bibitem{Xie:2013db}
  J.~J.~Xie and B.~C.~Liu,
  Phys.\ Rev.\ C {\bf 87}, 045210 (2013).

\bibitem{Xie:2014zga}
  J.~J.~Xie, J.~J.~Wu and B.~S.~Zou,
  Phys.\ Rev.\ C {\bf 90}, 055204 (2014).

\bibitem{Doring:2010ap}
  M.~Doring, C.~Hanhart, F.~Huang, S.~Krewald, U.-G.~Meissner and D.~Ronchen,
  Nucl.\ Phys.\ A {\bf 851}, 58 (2011).

\bibitem{Xie:2013wfa}
  J.~J.~Xie, B.~C.~Liu and C.~S.~An,
  Phys.\ Rev.\ C {\bf 88}, 015203 (2013).

\bibitem{Xie:2014kja}
  J.~J.~Xie, E.~Wang and B.~S.~Zou,
  Phys.\ Rev.\ C {\bf 90}, 025207 (2014).

\bibitem{Xiao:2015zja}
  L.~Y.~Xiao, Q.~F.~L\"u, J.~J.~Xie and X.~H.~Zhong,
  Eur.\ Phys.\ J.\ A {\bf 51}, 130 (2015).

\bibitem{Wang:2018vlv}
  A.~C.~Wang, W.~L.~Wang and F.~Huang,
  Phys.\ Rev.\ C {\bf 98}, 045209 (2018).

\bibitem{Wang:2017tpe}
  A.~C.~Wang, W.~L.~Wang, F.~Huang, H.~Haberzettl and K.~Nakayama,
  Phys.\ Rev.\ C {\bf 96}, 035206 (2017).

\bibitem{Huang:2012xj}
  F.~Huang, H.~Haberzettl and K.~Nakayama,
  Phys.\ Rev.\ C {\bf 87}, 054004 (2013).

\bibitem{Machleidt:1987hj}
  R.~Machleidt, K.~Holinde and C.~Elster,
  Phys.\ Rept.\  {\bf 149}, 1 (1987).

\bibitem{Liu:2006tf}
  B.~C.~Liu and B.~S.~Zou,
  Commun.\ Theor.\ Phys.\  {\bf 46}, 501 (2006).

\bibitem{Xie:2007qt}
  J.~J.~Xie, B.~S.~Zou and H.~C.~Chiang,
  Phys.\ Rev.\ C {\bf 77}, 015206 (2008).

\bibitem{Xie:2010yk}
  J.~J.~Xie and J.~Nieves,
  Phys.\ Rev.\ C {\bf 82}, 045205 (2010).

\bibitem{Xie:2015zga}
  J.~J.~Xie, Y.~B.~Dong and X.~Cao,
  Phys.\ Rev.\ D {\bf 92}, 034029 (2015).

\end{thebibliography}
\end{document}